## Interactive Employment Model to Assimilate the Deaf persons in workplace by using ICT

S. A. ALI[++] SAFEEULLAH, A. G. MEMON *, M. AHMED**

Department of Computer Science, Faculty of Information Technology, Sindh Madressatul Islam University, Karachi, Pakistan.

**Abstract**: The rate of disability is increase day by day all over the world. There are various type of Disabilities but the deaf persons are on second number among all types of disabilities.. In most of the countries disabled persons are supposed to be social liability on their family and in the society as awhile.

Now days in developing countries it is difficult for normal persons to get suitable jobs. Whenever we are talking about disabled, it is more difficult to assimilate deaf persons in workplace. Threats of unemployment of disabled person are almost double that of people without disabilities. Number of able deaf persons is unable to get suitable job due to several reasons like lack of facilities for deaf persons and lack of awareness from normal persons side which create barrier in searching job for deaf persons.

This research work emphasis on the special need and training required for the deaf individual sandto make and train them how to move in workplace with different social and technical barrier. In recent era technology play important role in each and every part of life. Using the facility of Information and communication Technology we can easily assimilate deaf in workplace. The proposed model helps deaf persons to adjust in their jobs.

**Keywords:** Interactive Employment Model, Deaf Persons.

### I. INTRODUCTION

The main purpose of this research is to induct the deaf persons in workplace by the use of Information and communication technology.

The Special persons having this disability lack in the problem of either hearing or speaking and sometimes both. The Dictionary defines **'deaf'** and **'deafness'** as wholly or partially without hearing and for most people who have no connection with deaf people this is a full enough description.

However, this definition does not indicate the problems that arise from deafness. Deafness is multifaceted.

*"A hearing impairment or hearing loss is a full or partial decrease in the ability to detect or understand sounds. Caused by a wide range of biological and environmental Factors, loss of hearing can happen to any organism that perceives sound "(*Oxford Dictionaries, 2010*).*

The modern technologies under the domain of education and Information technology can play an important role in promoting the education of Special persons. In modern education technologies, Computer technologies have become intrinsic and an important aspect of teaching and learning in a digital and networked society. Use of Artificial Intelligence, Information technology, theory of communication and multimedia can help special students in their education.

The use of technologies enhances the abilities of special students as well as provides support in their learning activities. Information and computer technologies has also enhanced the development of sophisticated devices that can assist a lot of students with more severe disabilities in overcoming a wide range of limitation that hinders classroom participation from speech and hearing impairments to Deafness and severe physical disabilities (Bryan, 2006).

Through the usage of ICT, it is comfort to design an **Interactive Employment Model** to communicate Information and Communication Technology for such special persons. Through the **Interactive Employment Model**, they can easily be adjusted in an encouraging atmosphere, which happens for normal person. To absorb and make them beneficial and active individuals in each and every stage of their lives, positive and independent solutions must be explored. In this research, an effort is being made to overcome the excessively high unemployment rate of deaf persons.

The Deaf are a distinct cultural-linguistic people group. This will be the biggest conceptual barrier you

---
[++] Corresponding author: Email: aasyed@smiu.edu.pk, ssoomro@smiu.edu.pk
*Institute of Mathematics and Computer Science, University of Sindh, Jamshoro
**Department of Computer Science, Benazir Bhutto Shaheed University, Karachi

face as you seek to understand the Deaf community and effective outreach within it. If you can step out of your hearing view of the world for a moment and see Deaf people as a cultural-linguistic group, then all that we do will begin to make sense. If you can't, and you continue to see deafness as simply a physical disability, then you probably won't understand or agree with much of what we say or do.

Historically and internationally, Deaf people live and interact as a distinct people group. They have their own customs, habits, thought patterns, language, common experiences, and values that identify them as a unique cultural group. They do not consider themselves handicapped or disabled; rather, they consider themselves minority group within the context of their own home country (Sternberg, 1998).

## 2. MATERIAL AND METHODS
**Joblessness rate of Disabled**

The joblessness rate between disabled people is double that of persons without disabilities.

Though they possess desirable or skills they have difficulties in finding job, or requisite at least work, adequate for their skills. Many of them have received training, but there are a few other functions where Deaf persons have worked effectively (Bryan Jones, 2004).

Efforts to adjust special persons into the workplace are not just a social compulsion. If deaf persons adjust, in job, they can increase their living standard and their families will grow, and the level of poverty will reduce.

Inopportunely, very few deaf persons though qualified, have problems in finding a suitable job. Even though there are a number of ways this problem may be solved, and one of them is to make sure whether they belong to any class in their lives like rehabilitation clients, students or working-age adults losing vision. They are ended to be mindful land aware of a handful of the numerous jobs that are being offered and could be successfully done in the work places.

**Jobs those Deaf Persons can do**

After proper training, these persons can do numerous jobs as they have like range of abilities that normal persons have except hearing and listening power. Therefore, the deaf person scan successfully accommodates them in the workplace.

There are numerous suitable and appropriate jobs that deaf persons can do are as follow (Majid, 2008), (David, Calder, 2009):

Assistant to manager, clerical assistant, stereotypes, typist, receptionist, computer operator, data entry operator, customer service provider, in showbiz etc.

Further, they can also serve as, inventors, factory workers; repair and service representatives, artisan, social worker shopkeepers etc (Country Profile on Disability, 2002).

**Barriers of employment for deaf Persons**

- One of the most universal and regular hindrance that is encountered by the deaf persons in their service is the attitude of their colleagues. This leads to their alienation in the workplace.

- This negative attitude of their employers or colleagues or other persons in society also erodes their confidence and self-image with eventual negative impact on their work performance (Kursha, 1959).

- Most of the employers have negative perception regarding the in ability and are more problematic than normal employees. This negative perception is due to lack of awareness among their colleagues and management.

- Deaf persons usually work without extra costs for their employer yet; they are discriminated even in terms of promotion as compared to normal employees.

- There is difference between academic training and practical work, they facecloth of problems, and even normal people face a lot the same problems.

- Further they have a separate cultural communication pattern with their own group called 'Sign Language'. This is the biggest barrier encountered in their work environment.

- When they join workplaces it poses real difficult for them to find bus, taxi etc. or move to the bus or taxi stand. In case of offices or business, transition is also very difficult for them to go to banks, withdraw funds and return safe and sound (J. KURSHA, 1959).

## 3. DISCUSSION AND RESULT
**Interactive Employee Model for the Deaf persons**

To design the interactive employee model using ICT following factors should be keeps in a mind.

i  Train Employer, How to behave with a Deaf employee?
ii  Reserve budget for providing assistive facilities to Deaf employees.

iii    Motivate and Encourage Deaf persons & fulfill their right.

iv    Internship to train and create awareness Deaf employees regarding work place.

v.    Provides approachable information, to search jobs.

To adjust the deaf persons into workplace first of awareness should be provide to normal persons how to deal with them.

People learn best by doing. During the training, this should be clear to normal employees and employers that, it is a social responsibility to show positive attitudes and cooperate with Deaf persons.

Due to lack of knowledge about ergonomics consideration, there is no proper seating arrangement for the deaf person in keeping with their comfort at work place (Khatoon,2003), ( Panayotova, 2007).

**How Technology Assimilate Deaf Persons into the Workplace**

To assimilate them in work organizations it is important to maintain database in order to identify the number of deaf persons in specific region or area .This database must be updated with the help of medical record. This statistical record helps government or NGO to assimilate deaf persons according to their qualification, experience, age group etc.

**Proposed Model to Assimilate Deaf Persons into the Workplace**

As deaf persons cannot read an application form in English, and are not able to search job vacancies and therefore, they cannot fill application forms for a specific job in an organization, even if that specific job requires low vision or no vision at all.

It is essential to be aware of their basic needs and demands. As deaf persons cannot listen and speak the first question how then go through interview process as they are not able to give face-to-face or telephonic interview due to their disability For their comfort, an interactive employee model is introduced to pass them through interview stage which is very first step for employment . With the aid of technology, deaf candidates can apply through email or can give interview through voice based internet software.

Adaptive technology becomes one of the vital tools in a case of enhancing the way to employment along with orientation and mobility guidance (Bryan, 2006).

In proposed system interviewer will ask questions to Interviewee in English, it will convert into sign language using the proposed model by

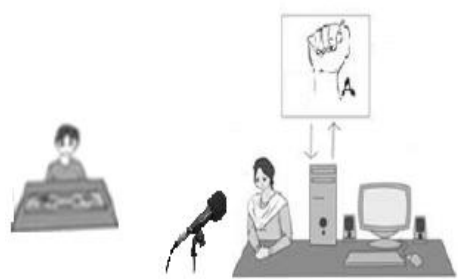

**Fig. 1**

Which deaf person can easily understand? If any things is not clear by interviewee, a Keyboard is provided to deaf candidate he/she can press 'R' key then a text message will speak that 'Please repeat your question"

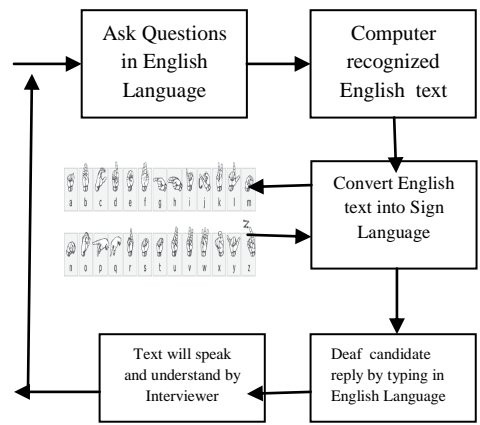

**Fig. 2**

The most important features of the proposed system mentioned below:

•    The Proposed System basically functions in order to provide a common platform practically for both type of candidates the normal and as well as the deaf candidate so that both groups of candidates will have the same sense of qualification achievement and may indeed achieve a comparable level of performance in the field of education.

•    The proposed System 'works in a very user friendly environment thus making the interviewer as well as the deaf candidates to feel at ease. There is no sort of any problem or obstruction in the flow and functioning of the proposed system.

- The interviewer does not need to learn the Sign Language (American Sign Language to educate the special (deaf) candidates.

- For the deaf candidates all possible interview questions can explain by their guider/parents which may asked my interviewer in English without learning Sign Language and the text will convert into sign language.

- To keep the deaf and even the normal people at ease an additional functionality of the facility of recording a particular lecture by their guider/parents is also available in the proposed system. Special students can easily listen to the same questions again and again as and when required without facing any difficulty.

There is no need to purchase expensive translators or devices to explain and make students understand special languages by their parents or guardians.

Using the same model deaf persons can give interview through cell phone is a very common technology for communication voice command cell phone which helps the deaf persons to minimum usage of device.

Mobile, wireless technologies such as mobiles and PDAs are drastically increasing as a new transfer source that provides high speed access in sending and receiving information, content, video voice. The latest and newly developed wireless technologies may allow mobile phones, and to be made transferable or computers that are wearable to perform their tasks as universal inaccessible consoles for obtaining data and facilities and control and monitor of different appliances and devices. To sum up, it can be said that wireless technology may soon become a vital and integral part in the lives of every individual and without its assistance the disabled may feel more neglected and ignored from the society and a number of activities (WHO, 2002).

## 5. CONCLUSION

It is a matter of fact and indeed, it is a reality that even today most of the employees with disabilities are hired as low-grade employees in the workplaces and organizations like peons, messengers, etc. Most of the hired employees are physically handicapped with polio problem and some are hearing impaired, very few were visually impaired and the number of mentally retarded is almost rejected.

This research provides a framework that makes it possible to integrate the deaf employees along with the normal employees in a way that is appropriate to the Deaf as well as the normal employees. This in turn leads to the enhancement of the selection criterion of the deaf employees in work environment.

This research also lays emphasis on the fact that the disability of deaf is neither a hindrance nor an obstruction in the flow of the employee selection for a particular job in a working organization and that employees even with this disability can apply for jobs in the same manner in which the normal employees.